\documentclass[letterpaper, twocolumn, 10pt]{article}
\usepackage{epsfig,endnotes}
\usepackage{booktabs}

\usepackage[utf8]{inputenc}
\usepackage[T1]{fontenc}
\usepackage{lmodern}
\usepackage[english]{babel}
\usepackage{epsfig,endnotes}
\usepackage{url}
\usepackage{pgfplots}
\usepackage{tikz}
\usepackage{multirow}
\usepackage{subcaption}
\usepackage[ruled,vlined]{algorithm2e}
\usepackage{amsmath}
\usepackage{graphicx}
\usepackage{caption}
\usepackage{array,ragged2e}
\usepackage{makecell}
\newcommand{\Xomit}[1]{}
\newcommand\remove[1]{}
\newcommand\ignore[1]{}
\usepackage{enumitem}
\usepackage{changes}
\usepackage{hyperref}

\usepackage{color}

\newcommand{\ronen}[1]{{\color{blue} #1}\normalcolor}

\begin{document}


\title{ Kubernetes Auto-Scaling: YoYo attack vulnerability and mitigation  }
\author{{\rm Ronen Ben-David}\\Interdisciplinary Center, Herzliya, Israel\\
ronen.bendavid@post.idc.ac.il
\and
{\rm Anat Bremler-Barr}\\
Interdisciplinary Center, Herzliya, Israel\\
bremler@idc.ac.il}

\date{February 25, 2021}

\maketitle



\begin{abstract}

In recent years, we have witnessed a new kind of DDoS attack, the burst attack\cite{ImpervaBurst,RadwareBlog}, where the attacker launches periodic bursts of traffic overload on online targets.  Recent work presents a new kind of Burst attack, the YoYo attack \cite{8057010} that operates against the auto-scaling mechanism of VMs in the cloud. The periodic bursts of traffic loads cause the auto-scaling mechanism to oscillate between scale-up and scale-down phases. The auto-scaling mechanism translates the flat DDoS attacks into Economic Denial of Sustainability attacks (EDoS), where the victim suffers from economic damage accrued by paying for extra resources required to process the traffic generated by the attacker. However, it was shown that YoYo attack also causes significant performance degradation since it takes time to scale-up VMs. \\
In this research, we analyze the resilience of Kubernetes auto-scaling  against YoYo attacks. As containerized cloud applications using Kubernetes gain popularity and replace VM-based architecture in recent years. We present experimental results on Google Cloud Platform, showing that even though the scale-up time  of containers is much lower than VM, Kubernetes is still vulnerable to the YoYo attack since VMs are still involved. Finally, we evaluate ML models that can accurately detect YoYo attack on a Kubernetes cluster.

\end{abstract}
\section{\uppercase{Introduction}}
\label{sec:introduction}
\noindent 
The burst attack is a new trend in DDoS attacks. In a burst attack, the victim is attacked with recurring high-volume traffic bursts, lasting from a few seconds up to several minutes, while the total duration of the attack can span hours and even days \cite{RadwareBlog,ImpervaPulse}. In 2017, it was reported that $42\%$ of organizations experienced such attacks \cite{bleeping_computer}.  Burst attacks have become feasible due to recent improvements in botnet technologies that allow them to be perpetrated in a scheduled and synchronized manner. A burst attack, when carried out correctly, is cost effective to the attacker, who can divert the attack traffic between multiple end-point targets, leveraging a high degree of resource control. Moreover, it confused conventional DDoS mitigation and detection solution \cite{ImpervaBurst}.

Recent work presents a new kind of Burst attack \cite{8057010}, the YoYo attack,  that operates against
 web services implemented as native cloud applications using VMs.  We note that cloud applications are resilient to many of the classic DDoS (i.e.,attackers operate a flat vector attacks) due to their high bandwidth pipe, built-in anti-spoofing mitigation in the load-balancers, and the common use of a content distribution network (CDN).  Amazon lists the auto-scaling mechanism such as using AWS layer DDoS mitigation as one of the best practices for dealing with Distributed Denial of Service (DDoS) \cite{amazonDDoSwhitepapers} attacks that target web applications. In flat DDoS attack the auto-scaling mechanism translates a DDoS attack into an Economic Denial of Sustainability attack (EDoS), incurred by paying the extra resources required to process the bogus traffic of the attack. However, there is no performance damage, since the extra machines handle the extra traffic.

The YoYo attack \cite{8057010} operates against the auto-scaling mechanism of VMs in the cloud. The periodic bursts of traffic loads cause the auto-scaling mechanism to oscillate between scale-up and scale-down phases. The YoYo attack causes significant performance degradation in addition to economic damage.  During the repetitive scale-up process, which usually takes up to a few minutes, the cloud service suffers from a substantial performance penalty.  When the scale-up process finishes, the attacker stops sending traffic and waits for the scale-down process to start. When the scale-down process ends, the attacker begins the attack again, and so on. Moreover, when the scale-up process ends, there are extra machines, but no extra traffic. Thus, the victim pays unwittingly for extra machines that are not utilized.

A recent trend in cloud applications is to use containers, adding an additional virtualization layer by loading multiple containers on a single VM. Containers are light-weight with lower space and time overhead than VMs. Thus, the question arises whether containerized applications are less vulnerable to the YoYo attack. One might speculate that this is the case due to their short scale-up and scale-down times while the performance penalty is proportional to scale-up time, and the economic penalty is proportional to scale-down time.\\ 
In this paper, we analyze Kubernetes, the most common container orchestration engine for containerized applications. We show that the YoYo attack can still have a huge impact. This is due to the Kubernetes cluster architecture, which combines two levels of virtualization, Pods (container level) and Nodes (VM level), and uses a two-stage auto-scaling mechanism: Pods are scaled first until they fill allocated nodes and trigger new node allocations. 
The root observation is that the increase in the number of nodes increases the economic damage of the attack but also the performance damage, since it takes time to scale up Node. In Section \ref{sec:Kubernetes} we present a model of Kubernetes auto-scaling mechanism.  We then present a formal model  of YoYo attack on Kubernetes in Section \ref{sec:formal}. 
In Section \ref{sec:evaluation} we evaluate YoYo attack on Google Cloud infrastructure. We compared the damage caused by the attack, economic and performance wise, while conducting YoYo attack on Google Cloud Engine (GCE) that is VMs based, and YoYo attack on Google Kubernetes Engine (GKE) that is containers based. We show, that Kubernetes, with the fast Pod scale-up reduces the performance damage, but does not eliminate it. We also evaluate the classic DDoS attack on Kubernetes and show that the YoYo attack is more cost effective to the attacker than a flat DDoS attack.

In Section \ref{sec:detection}, we propose a detection mechanism of YoYo attack on Kubernetes, based on machine-learning technique. Previous machine learning based techniques in the literature \cite{8748836,8460025} are not applicable to YoYo attack, since they rely mainly on traffic bandwidth and attempt to assign a score that reflects the severity of the attack. Our solution suggests the XGBoost classifier to detect a YoYo attack on a Kubernetes cluster. We show high accuracy results based on a comprehensive analysis using that model, exploiting unique data (i.e., response time, Pod count, Node count and CPU load) acquired from the Kubernetes cluster.

\section{\uppercase{Kubernetes}} \label{sec:Kubernetes} 
\subsection{Kubernetes Background} 
Kubernetes is an open-source container orchestration engine for automating the management and deployment of containerized applications. 
In this paper, we focus on Google Kubernetes Engine (GKE), but note that other implementations of Kubernetes in the cloud, such as AKS(Azure) and EKS(AWS), share the same behavior. We describe the Kubernetes basic mechanism, focusing on the aspects relevant to auto-scaling.
Kubernetes is a cluster of machines, called \emph{Nodes}, that are used to run user workloads called \emph{Pods}. 
A \emph{Pod} is the basic compute unit. It has one or more containers running together, according to specific configuration parameters such as network utilization, storage, CPU utilization, memory and container image name.  \emph{Nodes} are  virtual  machines in the cloud environment, and usually several Pods run on the same Node. 

Applications use a simple Kubernetes Controller called Deployment, which provides declarative updates for Pods. Each Pod gets its own IP address. In a Deployment where  Pods are created and destroyed dynamically, the set of Pods running at one moment might differ from the set of Pods running that application a moment later. However, when there is a requirement to track Pods and their IP address, Kubernetes defines a Service, which is a logical set of Pods and a policy (aka microservice). The set of Pods targeted by a Service is determined by a selector. When a Node is closed, all the Pods that run in the context on the Nodes are also closed. 
Pods can run in standalone configuration but are usually grouped by Controllers. The Controller is responsible for maintaining the correct number of Pods for every replication. For simplicity, we demonstrate the impact of the YoYo attack on a deployment object and a deployment controller. 
Deployment objects represent a set of multiple, identical Pods (that can be parts of multiple Nodes).  Deployments are well-suited for stateless applications. We note that there are other types of objects, such as StatefulSets, which are designed for stateful applications. \ignore{ Namespaces in Kubernetes provide a way to divide the cluster into different logical clusters. These logical clusters are allocated to the specific teams or departments in an organization. When starting to analyze the Kubernetes resources, namespaces are a good starting point for cost analysis.} 
\subsection{Kubernetes Autoscaling}
\label{sec:cloud_auotscaling}
The GKE autoscaling is done by two components, the Horizontal Pod Autoscaler that automatically scales the number of Pods and works on the application abstraction layer, and the Cluster Autoscaler that automatically resizes the number of Nodes in a given Node pool and works on the infrastructure layer. Kubernetes also defines the Vertical Pod Autoscaler (VPA) that automatically sets the resource request and limit values of containers based on usage. VPA is not in the scope of this work.  
\subsubsection{The Horizontal Pod Autoscaler (HPA)} 
The HPA is responsible for automatically scaling the number of Pods \cite{HPAAutoscaling}. The controller periodically adjusts the number of Pods needed based on the observed metric to the target configured by the user. Each scale rule is defined by a threshold, scale interval and action, s.t.  if the threshold exceeds the duration of the scale interval, the action will be performed.  We denote by $I^p_{up} \backslash I^p_{down}$ the scale interval for scale-up and scale-down of a Pod. Note that the default values (correct to the time of writing this paper) are 1 minute for $I^p_{up}$ and  5 minutes for $I^p_{down}$ . The most common metric for a threshold is the relative CPU utilization, measured as the actual CPU usage of a Pod divided by the CPU requested by the Pod. Note that the different metrics are measured on a service level.
Let $P$ be the number of Pods, let $U_{target}$ be the relative target CPU of the Pod, defined as the recent CPU divided by the CPU requested by the Pod,  and let $U_i$ be the relative CPU utilization of the Pod $i$ measured across the last $1$ minute.
Note that the relative CPU utilization can have a value higher than $100\%$, since the Pod  CPU usage is configured in milli CPU units. Thus, 200 milli CPU is equal to 0.2 CPU, and if in peak time the Pod uses 500 milli CPU, then $U_i=250\%$.
We define the \emph{Average Relative CPU Utilization} (Current CPU Utilization Percentage in Kubernetes terminology \cite{KubernetesCodeReview}) as: \begin{eqnarray}
 \frac{\sum_{1 \leq i \leq P}{U_i}}{P} 
\label {eq:hpa_average}
\end{eqnarray}
Note that, similar to relative CPU utilization, the average CPU utilization can be higher than $100\%$.
The goal of the HPA is that the value will be close to the target value, in our case CPU utilization,  $U_{target}$. In order to avoid oscillation, the HPA triggers scaling actions only if the value is below 0.9  or above 1.1 of the target value (i.e., 10\% tolerance) \cite{HPAAutoscaling}. Thus, the target number of Pods is:
\begin{eqnarray}
\lceil \frac{\sum_{1 \leq i \leq P}{U_i}}{U_{target}} \rceil 
\label {eq:hpa_pods}
\end{eqnarray}

\ignore{Daniel, why not using the HPA formulation? -> desiredPods = ceil[currentPods * ( currentMetricValue / desiredMetricValue )]}

We note that other possible metrics are the relative memory and storage. The HPA uses an internal service called a \emph{metric server} to periodically test the metrics of the cluster and act accordingly (scale-up/scale-down) \cite{HPAAutoscaling}. After a scaling decision is made, it takes relatively little time until the Pod is ready to function.  We call this time the \emph{Warming time} and we denote it by $W^p_{up}$. Similarly, we also have $W^p_{down}$, the time until the Pod is destroyed. We observed very fast warming time, less than $30$ seconds, and downtime of $5$ seconds.  

\subsubsection{The Cluster Autoscaler (CA)} The CA interacts with the HPA and the metric server. It monitors and populates pending (newly created) Pods and releases idle Nodes after the Pods are closed. Specifically, the CA checks every 10 seconds for pending Pods that cannot be allocated in existing Nodes due to the lack of computer resources. If such Pods are found, it initiates the creation of new Nodes to which the pending Pods are scheduled. The number of Pods in the Nodes is according to the machine type of the Nodes and the deployment configuration \cite{ClusterAutoscaling}. 
\ignore{Daniel, and the deployment configuration. e.g. deployment that requires 300vmcpu per Pod could have up to 3 Pods per Node with 1 vCPU usualy less due to the Pods the control plane needs].} 
The machine type is configured by the system administrator (per Node pool), and the system allocates as many Pods as possible to a Node. The bound on the minimum and maximum number of Nodes in a cluster can be configured. A minimal Kubernetes cluster needs 3 Nodes for scalability and high availability and can scale up to thousands of Nodes \cite{scalability}. We denote by the $I^n_{up} \backslash I^n_{down}$  the scale interval for scale-up and scale-down of Nodes and the warming time of Nodes by $W^n_{up}$ and $W^n_{down}$. We observed interval scale-up of 10 seconds, and interval scale-down of 10 minutes, which corresponded to a scale-up warming time of 2 minutes and scale-down warming time of around $2$ minutes.  We note that none of these parameters can be configured and are set by the Kubernetes infrastructure.

\begin{figure*}[!h]
  \centering
  {\epsfig{file=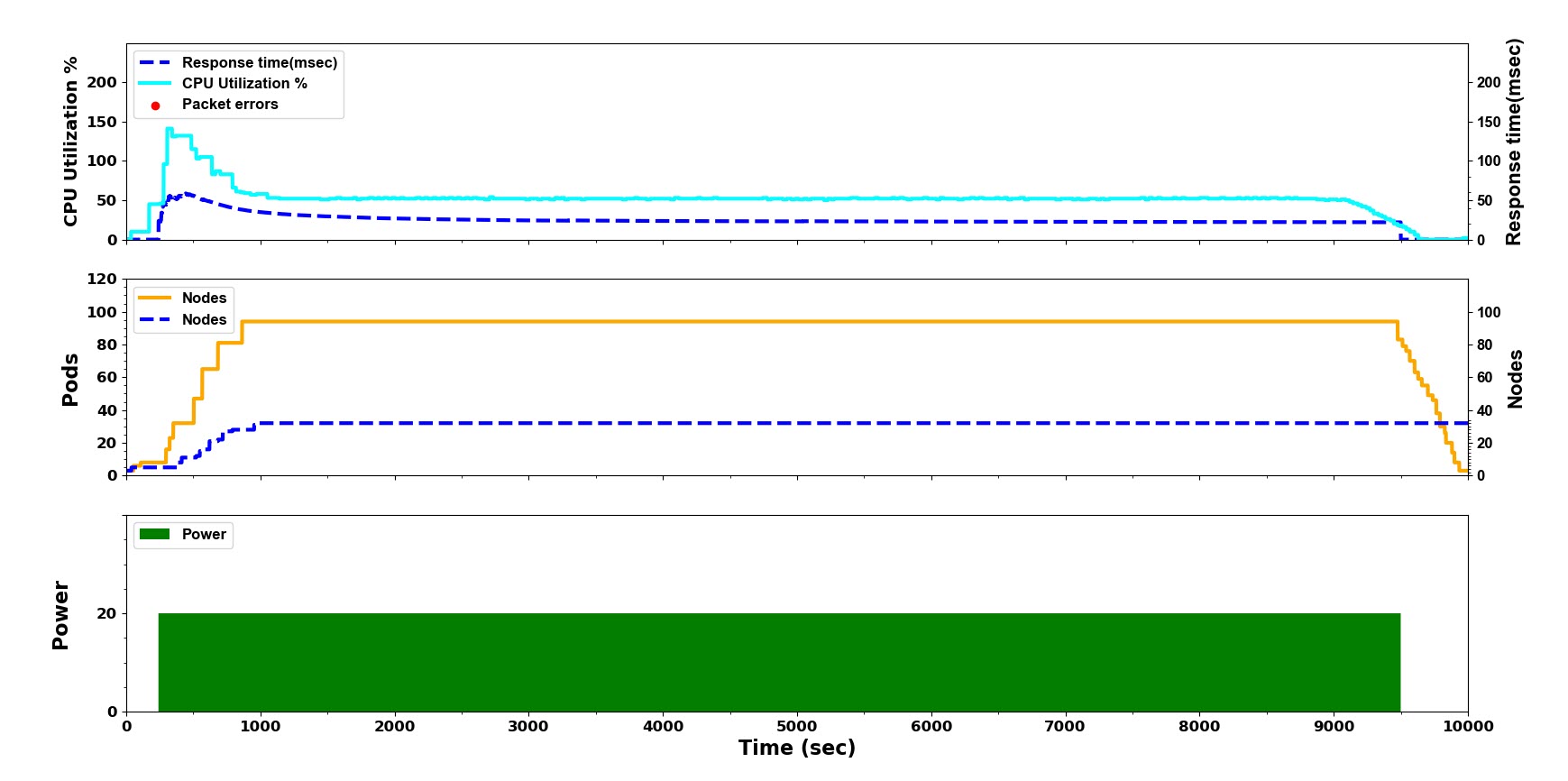,height=3.05in}}
  \caption{Classic DDoS attack with power k=20 on Kubernetes.}
  \label{figure:classic_attack}
\end{figure*}

The Kubernetes autoscaling mechanism is well illustrated in figures \ref{figure:classic_attack}. It shows nicely how the cluster scales-up Nodes and Pods to manage a flat DDoS attack.
 
\subsection{Kubernetes Pricing Models}
As Kubernetes has become the leading container orchestration tool in the market, major Cloud vendors have developed different pricing models to support their customers in order to leverage Kubernetes from the TCO (total cost of ownership) perspective.  Cost analysis is crucial to understanding the economic damage of a YoYo attack on a Kubernetes cluster. We have selected Google Kubernetes engine platform as our choice to analyze YoYo attack experiments. However, our analysis and conclusions are relevant to Kubernetes technology in general.  The cost of running a Kubernetes Cluster in GKE is mainly a function of the number of Node instances. Customers are billed for each of those instances according to the Compute Engine's pricing, until the Nodes are deleted. These Node instances are billed on a per-second basis with a one-minute minimum usage cost regardless of the container workload in that instance.  In addition, there is an hourly flat fee for cluster management, irrespective of cluster size and topology, whether it is a single-zone, multi-zone or a regional cluster. Amazon EKS has a similar pricing model where customers using Amazon EC2 will pay for EC2 Node instances. Amazon established an alternative pricing model for their customers called AWS Faragate where the customers are charged per vCPU workload, meaning per resource \cite{amazonEKSPricing}. 

In a Faragate Kubernetes cluster, the instance hours are proportional to the number of Pods in the cluster and other resources allocated to those Pods.  Therefore, the cost of attack is derived from the performance damage accrued, and the load of requests that the cluster can process concurrently is influenced by the number of Pods and their elasticity.
\section{\uppercase{YoYo attack }}
\label{sec:formal}

\begin{figure*}[!htbp]
\centering
\psfig{file=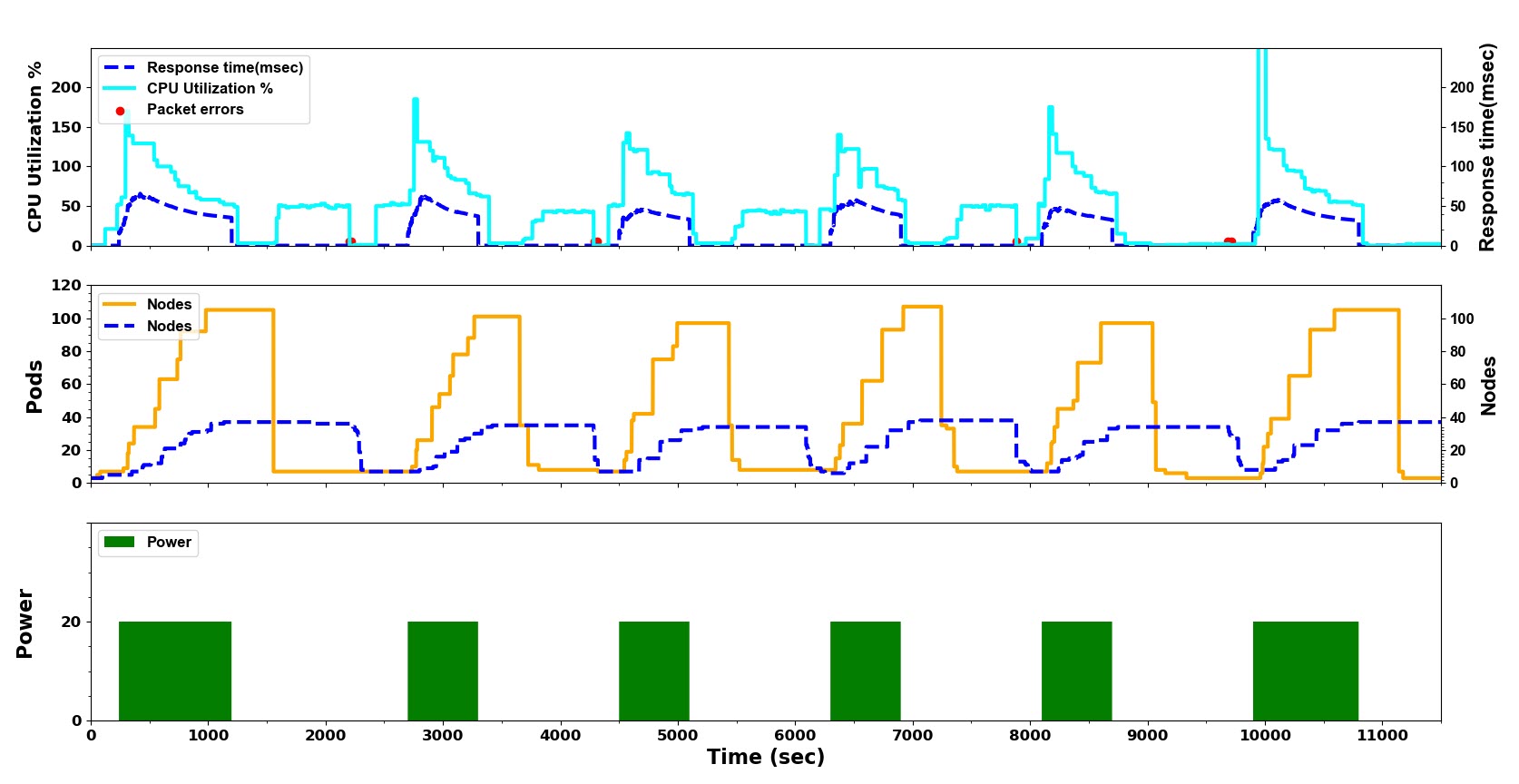,height=3.05in}
\caption{ YoYo attack with power k=20 on Kubernetes }
\label{figure:yoyo_attack} 
\vskip -6pt
\end{figure*}

We follow the notation of the original YoYo attack \cite{8057010} and adapt it to the Kubernetes model. Consider a Kubernetes environment that includes autoscaling with identical service machines behind a load balancer.  Requests arrive with average rate of $r$ requests per unit time,  and the load balancer distributes them to $N_p$ Pods that are divided between $N_n$ Nodes in the steady state. Let $R$ be the number of Pods that can be in one node. The number of Pods in a Node is determined by the Node machine type; a stronger machine will have a higher number of Pods. 

The YoYo attack is composed of $n$ cycles, and each cycle duration is $T$, comprised of an \emph{on-attack} period, denoted as $t_{on}$, and an \emph{off-attack} period, denoted as $t_{off}$. Thus $T = t_{on} + t_{off}$. We define the \emph{power of the attack} as the extra load on the cluster. Let $k$ be the power of the attack and $r$ the average request rate per second in a steady state. We assume that in the \emph{on-attack} period, the attacker adds fake requests $k$ times more than the rate in the steady state (i.e. a total rate of $(k+1) \cdot r$), while in the \emph{off-attack} period $t_{off}$, the attacker does not send any traffic. See Table \ref{tbl:notation}  for notation summary.
The following is the best strategy from the adversary side: 
We assume that the attacker aims to optimize the economic damage of the attack, with the main goal of being active as little as possible while still scaling up to extra $k\cdot N_n$ nodes. A secondary goal is to cause performance damage. 

Autoscaling will occur as a result of Pod creation, which automatically activates  the creation of new Nodes. Two conditions must be met in order to activate autoscaling: First, the extra load of $k \cdot r$ should burden the system such that the threshold for Pod scaling is fulfilled, regardless of the criterion (e.g., CPU utilization, traffic). Second, $t_{on}$ should be greater than or equal to the scale-interval of the Pods: $I^p_{up}$. That is, $t_{on}\geq I^p_{up}$. 
To maximize the performance damage, the value of $t_{on}$ should be set such that the attack continues the load up until all Pods are active and it loads the largest possible number of Nodes set in the cluster. Node loading occurs throughout $t_{on}$ since there are not enough activated Pods to meet the performance criterion.\\ That is, $t_{on}=I^p_{up}+W^p_{up}+I^n_{up}+W^n_{up}$.

Note that the dominant values are $I^p_{up}$, which is around $1$ minute, and $W^n_{up}$, which is around $2$ minutes. Therefore an optimal $t_{on}$ is around $4$ minutes. Moreover, most of the parameters use the system default configuration, but some of them can be modified. As such, the attacker will know most of these parameters and can take advantage of them to optimize the attack. The $t_{off}$ should be large enough such that all the Pods and Nodes perform scale-down as this is how we maximize the run time of the Nodes and cause extra spending.
The mechanism first scales down all Pods, after which it triggers the scale-down of the Nodes. Hence, $ t_{off}=I^p_{down}+W^p_{down}+I^p_{down}+W^n_{down}$. Note that the dominant values are $I^p_{down}$ of $5$ minutes and $I^n_{down}$ of $10$ minutes. Thus, the optimal $t_{off}$ is around $18$ minutes. 
\ignore{
{We define $D_p^{attack}(k)$, the performance damage caused by an $attack$ as a function of $k$, the power of the attack, and assess it as the average extra response time to answer requests during the total attack time. We note that this is a simplified assumption, since \ignore{due to network protocols (such as TCP and HTTP),} the actual impact on client performance is more complicated to analyze. We define the relative performance damage as the ratio between the damage following the attack and the corresponding value measured at a steady state.
\begin{eqnarray}\label{eq:rpd} RD_p^{attack}(k)= \frac{D_p^{attack}(k)}{D_p^{attack}(k=1)}.\end{eqnarray}
Similarly, we define $D_e^{attack}(k)$, the economic damage caused by the attack, and assess it as the average extra Nodes running in the system for the duration of the attack:
}
\ignore{Let us define relative damage as the ratio between the damage following the attack and the corresponding value measured at a steady state. We denote $RD_{p}^{attack}(k)$, $RD_e^{attack}(k)$  as the relative performance damage and relative economic damage correspondingly.} We refer in Figure \ref{figure:potency} to the following expression: \begin{eqnarray}\label{eq:red} RD_e^{attack}(k)= \frac{D_e^{attack}(k)}{D_e^{attack}(k=1)}.\end{eqnarray}}

An illustration of the YoYo attack can be seen in figures \ref{figure:yoyo_attack}. It shows nicely how the cluster oscillates its auto-scaling mechanism bases on Pods and Nodes to manage bursts of escalated traffic. 
\begin {table*}[ht]
\caption{Notation used in the model description. The values of parameters given according to the experiment  in Section \ref{sec:evaluation}.}
\vskip 2pt
\label {tbl:notation}
\centering
\resizebox{\linewidth}{!}
{
\begin{tabular}{|l||l|l|l| }
\hline
Parameter & Definition & Configuration given by & Value  \\
\hline
\hline
$r$ & Average requests rate per second of legitimate clients & System usage & \\
\hline
$N_p$ & Initial number of Pods  &  & 4 \\
$N_n$ & Initial number of Nodes & System administrator  & 4 \\
$R$ & Number of Pods per Node & & 3   \\
$I^{p}_{up} \backslash I^{p}_{down}   $  & Threshold interval for scale-up and scale-down for a Pod & &  1min$\backslash$5min\\
\hline
 $I^{n}_{up} \backslash I^{n}_{down}$  & Threshold interval for scale-up and scale-down for a Node &   & 10sec$\backslash$10min \\
$W^{p}_{up}  \backslash W^{p}_{down}$ & Warming time of scale-up and scale-down for a Pod & Kubernetes infrastructure & 30sec$\backslash$5sec \\ 
$W^{n}_{up}\backslash W^{n}_{down}$ & Warming time of scale-up and scale-down for a Node &  & 2min$\backslash$2min \\
\hline
$k$ & The power of the attack  & & \\
$n$ & Number of  attack cycles & Attacker  &\\
$T$ & Cycle duration &  & 10$\backslash$20 \\
$t_{on}  \backslash t_{off}$ & Time of \emph{on-attack} phase and  \emph{off-attack} phase. $T=t_{on}+t_{off}$& & \\
\hline
\end{tabular}}
\end {table*}
\section{\uppercase{Experiment Analysis}} \label{sec:evaluation}
In this section, we present a proof of concept of the YoYo attack on Google Kubernetes engine (GKE). Our environment in GKE consists of a simple HTTP server wrapped with a Docker container, front-end side stateless without back-end.  Each container is wrapped into a Pod.  We ran the Pods with the Deployment Controller, which is well-suited for stateless applications.\\

\subsection{GKE Parameter Settings}
We set the HPA utilization target parameter, $U_{target}$, to $50\%$. The container runs a Web server where each connection requires high CPU consumption. Each request to the Web server will perform some computation on the dynamic input of that query. We used \emph{Google Stackdriver} and the Kubernetes "watch" command to monitor cluster parameters while collecting logs using the Kubernetes Core API about Pods, Nodes, Relative CPU Utilization and the response time.  We used Apache JMeter \cite{Jmeter} to simulate the load on the cluster. We evaluated the performance and economic damage throughout the attack. Our  Cluster Autoscaler boundaries were set to a minimum of $3$ Nodes. 
For the Node machine, we use an N1-Standard-1 CPU, and we started the experiment with $4$ Nodes. In this configuration there are 3 Pods per Node ($R=3$). At the beginning of the experiment there are 4 Nodes and 3 Pods ($N_n=4$ , $N_p=3$). Hence, the system has enough Nodes to populate newly created Pods (at least more than $N_n*R-N_p$ Pods). This is a common technique that allows a fast response to a limited overload, while paying for extra Nodes.
We set the \emph{power of the attack} k, to $20$. We set the on-attack time to $t_{on}=10$ minutes and the off attack time to $t_{off}=20$ minutes; a substantial number of experiments showed these values to be the best (although not optimal bullet proof) for the YoYo attack and can demonstrate YoYo attack characteristics.
\begin{figure*}[!ht]
  \centering
 {\epsfig{file = 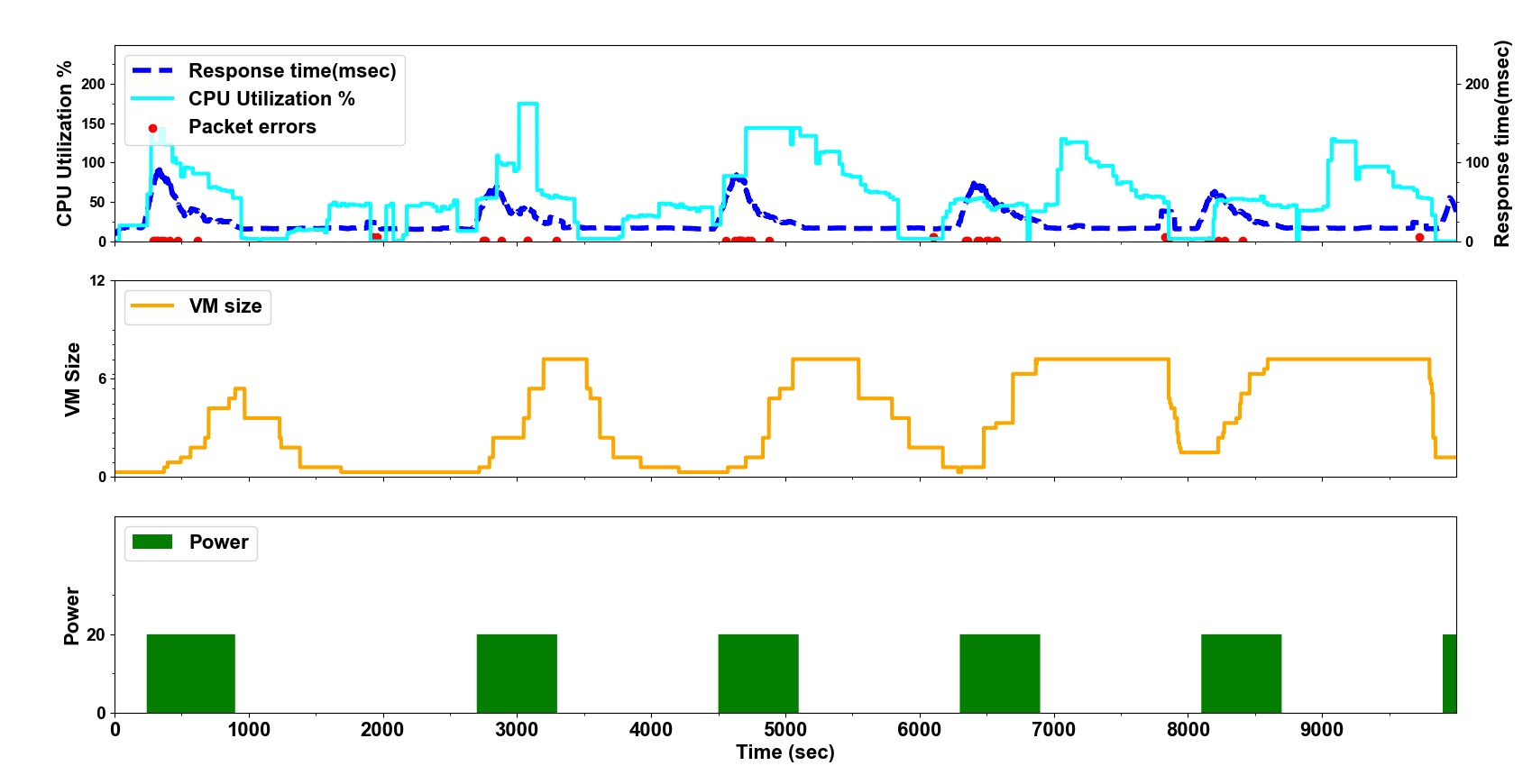, height=3.05in}}
  \caption{YoYo attack on Google Cloud Engine, VM Group.}
  \label{figure:vmg}
  \vspace{0.8cm}
\end{figure*}



\ignore{Figure \ref{figure:yoyo_attack} shows average CPU target in percentage  ,number of Pods , number of nodes and the response time in milliseconds under cluster autoscaling with the power of 20 on a cluster of 36 nodes, each node is configured with machine type of  n1-standard-1 (low cost). The figure clearly shows economic damage caused by the increase in the number of nodes, as predicted by the model. Furthermore, we were confidence to claim that the response time graph in the figure perfectly matches to our initial model prediction as the increase of Pods observes most of the attack when power of the attack is within a reasonable scope. We represent in a single graph the http error instances, the response time incidents exceeding 200 msec and the a mean of response time  across a sliding window of 60 seconds through the cycles of the attack.}

\subsection{Experiment Results}
Figure \ref{figure:yoyo_attack} illustrates this attack:  There are three sub-graphs sharing the x-axis, that shows the attack time series in seconds. The bottom sub-graph shows the $t_{on}/t_{off}$ attacks (on attacks are in filled green rectangles) using the \emph{Power} unit. The middle sub-graph describes the expansion of Pods (orange line) that the victim loads throughout the attack. The dotted blue line represents the number of nodes provisioned to accommodate the Pods. The number of Pods increases up to 120 and the number of Nodes increases to 32 as a result of the increase in CPU utilization. When the attack ends the cluster autoscaling keeps Pods in the idle state for 5 more minutes and the Nodes remain active for 10 additional minutes after all Pods are terminated. 
The top sub-graph describes both the average relative CPU utilization in cyan solid line. The CPU Utilization graph follows the equation \ref{eq:hpa_average}, and as explained, the value can be larger than 100\%. The HPA aims for a value of $U_{target}=50$, and the scaling decision is made according to these values.  The dashed blue line in the sub-graph shows an average response time to answer requests in the cluster.  \\
In some cases when nodes are deleted we experienced transient disruption interpreted as packet errors and marked in red dots. This phenomenon is due to the fact that our workload consists of a controller with a single replica, whose Pod might be rescheduled to a different node when its current node was deleted. This can be solved by configuring critical Pods not to be interrupted.

\subsubsection{Comparing the YoYo attack to the Classic attack}
In order to understand the effect of the YoYo attack on Kubernetes as compared to a traditional DDoS attack, we executed a  constant power attack experiment ($k = 20$). Figure \ref{figure:classic_attack} illustrates a classic DDoS attack that employs the same power of attack on the cluster as in the YoYo attack. The three sub-graphs share a common x-axis which, as in Figure \ref{figure:yoyo_attack}, shows the attack time series in seconds. Likewise, the y-axis of each sub-graph is as described in Figure \ref{figure:yoyo_attack}. 

We define $D_p^{attack}(k)$, the performance damage caused by an \emph{attack} as a function of $k$, the power of the attack, and assess it as the average extra response time to answer requests during the total attack time. We note that this is a simplified assumption, since the actual impact on client performance is more complicated to analyze. We define the relative performance damage as the ratio between the damage following the attack and the corresponding value measured at a steady state.
\begin{eqnarray}\label{eq:rpd} RD_p^{attack}(k)= \frac{D_p^{attack}(k)}{D_p^{attack}(k=1)}.\end{eqnarray}
Similarly, we define relative economic damage as the ratio between the economic damage following the attack and the corresponding value measured at a steady state. \begin{eqnarray}\label{eq:red} RD_e^{attack}(k)= \frac{D_e^{attack}(k)}{D_e^{attack}(k=1)}.\end{eqnarray} 
$D_e^{attack}(k)$, is the economic damage caused by the attack, and assess it as the average extra Nodes running in the system for the duration of the attack.
We refer in Figure \ref{figure:potency} to $RD_{p}^{attack}(k)$ and $RD_e^{attack}(k)$ as the relative performance damage and relative economic damage correspondingly. 

In the YoYo attack, the attack cost is directly affected by the power of attack $k$ and by the $t_{on}$ period relative to the attack cycle length. That is: \begin{eqnarray}\label{eq:cost} Cost(k)=k\cdot \frac{t_{on}}{T}\end{eqnarray}
The cost of a classic DDoS attack is equal to the power of the attack, while the cost of the YoYo attack is only a third of the cost of a classic attack.  

Potency is a metric that measures the effectiveness of DDoS attack from the attacker side. It is defined as the ratio between the damage caused by an attack and the cost of mounting such an attack. We denote $P_e(k)$ as the Potency of the attack.  An attacker would be interested in maximizing the potency. We use the following definitions for cluster autoscaling (ca) attack:
\begin{eqnarray}\label{eq:potency}
P_e(k) = \frac{RD_e(k)}{Cost(k)}\end{eqnarray}
Figure \ref{figure:potency} illustrates a comparison of the relative economic damage and the potency incurred by the victim had it been attacked by the two attack types. Figure \ref{figure:potency} shows that the classic DDoS attack results in $RD_e=7$ whereas the YoYo attack results in $RD_e=5$. The top sub-graph shows the potency comparison between the YoYo attack and the classic DDoS attack. In addition, the YoYo attack causes iterative performance damages, incurred at the beginning of each iteration starting at $t_{on}$. The classic DDoS attack causes the least performance damage to a Kubernetes cluster without any packet errors since Pods are not rescheduled and Nodes are not scaled-down.  The attack impact is mainly in the beginning of the attack while for the remainder of the attack a Kubernetes cluster is fully resilient. 
We summarize the results in Table \ref{tbl:compare} and conclude that the YoYo attack on Kubernetes is more cost effective for the attacker than a classic DDoS attack.
\subsubsection{YoYo attack: Kubernetes Vs. VM}
\begin{figure*}[!ht]
\centering
\psfig{file=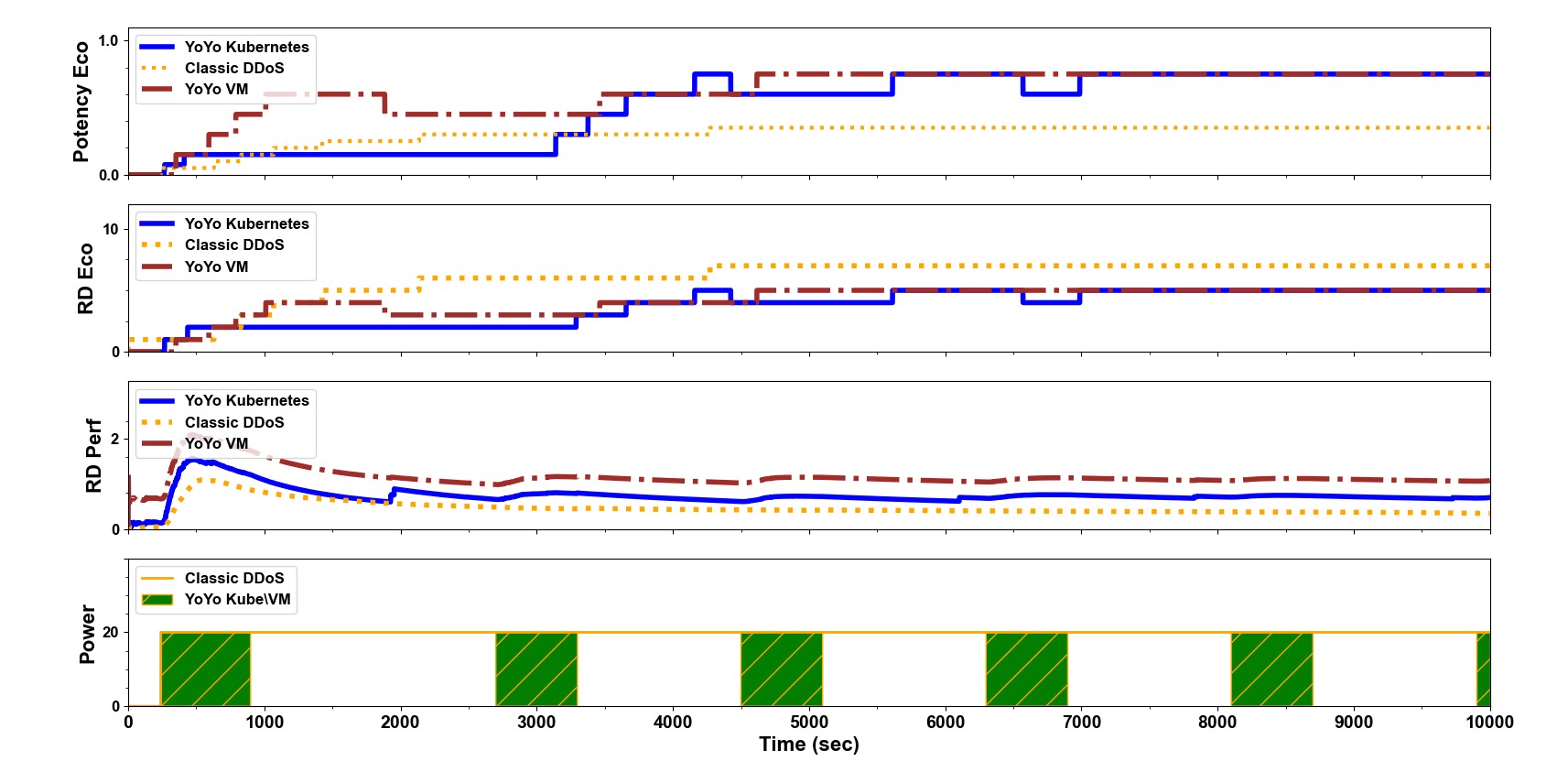,height=3.05 in}
\caption{Measuring attack effectiveness: Classic DDoS Vs. YoYo on Kubernetes and VM.}
\label{figure:potency} 
\vskip -6pt
\end{figure*}

To emphasize that Kubernetes has better resilience than VM against YoYo attacks (performance wise) but shares a similar vulnerability to economic damage, we repeated the experiments from the original YoYo paper \cite{8057010} and compared them to the experiments we conducted in this paper. We built an instance of VM group in GCE and a load balancer based on a VM group template using machine type \emph{n1-standard-1 (1 vCPU, 3.75 GB memory)} identical to the one we built for the GKE cluster.  The VM group instance is set with auto-scaling capability of up to 24 VMs and  adaptive scaling where all the machines that are ready to be scaled are scaled up at once. 
We used same parameters used in the YoYo Kubernetes attack and ran it for n cycles of duration T.  The power of the attack was $k=20$. {Figure \ref{figure:vmg} illustrates YoYo VM attack which can be compared with the YoYo Kubernetes attack illustrated in Figure \ref{figure:yoyo_attack}. A key observation is that an attack on VM group causes to an immediate slow response time and packet errors through the burst of loads. That observation lasts until the scale-up process ends. This behavior repeats in every attack cycle. YoYo VM attack results a relative performance damage $RD_p=1.66$. The relative performance degradation recorded by a YoYo Kubernetes is significantly lower with almost no packet errors (except the transient disruption due to Pods rescheduling). We can explain this since Kubernetes loads Pods in seconds to absorb the increased traffic until enough additional Nodes are ready with new Pods. Like in Kubernetes the attack cost is directly affected by the power of attack $k$ and by the $t_{on}$ period relative to the attack cycle length. Figure \ref{figure:potency} shows that the YoYo VM attack results almost the same relative economic damage and Potency as the YoYo attack causes to a Kubernetes cluster. YoYo attack may cause relatively the same economic damage in the cloud to GCE VM group as to a Kubernetes cluster, while the performance damage is more significant to a victim in GCE VM group. We conclude that a VM group is less resilient to YoYo attacks than a Kubernetes cluster.
\ronen{
\begin {table}[ht]
\caption{Summary results: Classic DDoS:$(k=20)$, YoYo Kubernetes and YoYo VM both:$[{t_{on}=10,t_{off}=20]}$,$(k=20)$} \label{tbl:compare} \centering
\resizebox{\linewidth}{!}
{
\begin{tabular}{|l||c|c|c|}

\hline
Parameters & Classic DDoS & YoYo Kubernetes & YoYo VM \\    
\hline
\hline
$Cost$ & $20$ & $\frac{20}{3}$ & $\frac{20}{3}$\\
\hline
Relative Economic Damage  & $7$ &  $5$  &  $5$\\
\hline
Relative Performance Damage & $0.75$  & $1.15$  & $1.66$\\  
\hline
Potency Economic  & $0.3$ &  $0.75$ &  $0.75$\\
\hline
\end{tabular}}

\end {table} }}




\ignore{Figure \ref{figure:yoyo_attack} combines Yo-Yo (\ref{fig:yoyo_attack}) and constant DDoS attacks (\ref{fig:classic_attack}) with the same power on the same cluster.{ The left set of graphs describe the YoYo attack and the right set of graphs describe a typical DDoS attack.} We refer mainly to the Yo-Yo attack although the figures share the same parameters and have the following attributes:
X axis describes  time series in hours:minutes. Y axis has multiple graphs, all are synchronized against the attack time series. Here we give explanation on the graphs from bottom graph to up. Bottom graph shows the $t_{on}/t_{off}$ attacks (on attacks are colored in green) using \emph{Power} unit. 
{ In each attack, T is the duration of each cycle. We then show the Total CPU load  which is the numerator in equation \ref{eq:hpa_pods} grows up to a certain value and increases the number of Pods in gradual steps to a level that TCL decrease below $U_{target}$, our desired CPU utilization.}

Next graph we show the average relative CPU utilization (according to Equation 
 \ref{eq:hpa_average}, and as explained the value can be larger than 100\%). The HPA aims that the value will  be equal to $U_{target}=50$, and according to these values the scaling decision is taken. 
 
 Next graphs are the number of Nodes and Number of Pods as function of time.  When the attack has ended it is clear that the Pods remains for 5 minutes, and then the Nods remain 10 minutes after Pods are terminated. 
 
Next graph, is the response time. The performance damage is the extra response time victim's suffers because of the attack. On the beginning of each attack phase  we observe  a significant increase in response time of around 100 msec. Eventually Kubernetes after scaling maintains backs the low response time (below 25 msec).
Note that in addition there are correlative disruptions to the scale down of the Nodes.   A significant increase in response time (200-1200 msec). We note that we also observe few Http errors correlated with the increase in response time.  The phenomena was also mentioned at  \cite{KubernetesCodeReview}. 

Top graph,  we show the \emph{ Relative Economic Damage} which is defined as follow: We define  the economic damage caused by the attack and assess it as the average absolute \textbf{extra} Nodes running in the system during the total attack time.
I.e., the patterned color area in the Nodes graph above $N_n$ represents the economic damage to the victim. Relative damage is defined as the ratio between the damage following the attack and the corresponding value at steady state, which in our case is closed to $5$.
In order to understand the effect of Yo-Yo attack on Kubernetes comparing to traditional DDoS attack, we did an experiment where we attacked with DDoS all the time with power of $20$ (see Figure \ref{fig:f2}). We give the summary of the result in Table \ref{tbl:compare}. 
In Yo-Yo attack, the attack cost is directly affected by the power of attack $k$ and $t_{on}$ period relative to attack cycle length. $Cost(k)=k\cdot \frac{t_{on}}{T}$.

While the cost of the attack in DDoS is equal to the Power of attack, in our Yo-Yo attack it was a third of the cost. However, the Relative Economic Damage was closed (7 for DDoS and 5 for Yo-Yo attack), and in Yo-Yo attack there multiple performance damage, in the beginning of each $t_{on}$ phase and $t_{off}$ phase, while DDoS attack there was only one  performance damage, in the beginning of the attack. We conclude that Yo-Yo attack is more cost-effective for the attacker.
}
\ignore{ 
Economical Damage will be as follow:\\
\begin{eqnarray}\label{eq:ca_demage_e   }
{D}_e^{ca}= \frac{\sum_{i \in T}{(N_i - N_n)}}{T} 
\end{eqnarray} 
{From \cite{Kubernetes} and observations $W^n_{up}+I^n_{up} + W^n_{down}+I^n_{down}$  is approximately 15 minutes. Scaling up/down the Nodes occurs mainly discretely although in some cases multiple Nodes are added or destroyed at the same time. We can see at figure \ref{figure:ca_attack} that the victim of an attack k=20 pay a mean value of 18 extra Nodes for the attack duration. For simplicity we spare the calculation and derived that in this case  ${D}_e^{ca}= \frac{5\cdot{(k+1)}}{6}$}
}

\ignore{Here is the theory, reinforced with figure \ref{figure:ca_attack}: In effective attack at time $t_on$ Nodes climb for 4 minutes to $(k+1)\cdot m$ Nodes and remain for $I^n_{down}=10 minutes$ through the period $t_{off}$. Then it downscale back to $N_n$ for $W^n_{down}$.  Eventually the extra Nodes are break to $\frac{(k+1)}{2}$ through $\frac{T}{3}$ (discreet upscale Nodes policy) and cost a flat cost of $(k+1)\cdot m$ for $\frac{2\cdot(T)}{3}$.  Thus, the total cost equals to $(K+1)\cdot(\frac{2}{3}) + (K+1)\cdot(\frac{1}{6}) = (K+1)\cdot(\frac{5}{6})$. The Total Cost of extra Nodes equals to ${D}_e^{ca}(k) = \frac{5}{6\cdot(T)}\cdot(k+1)$. }

\section{\uppercase{YoYo Attack Detection}} \label{sec:detection}
In this section we propose an enhanced detection technique designed to protect Kubernetes autoscaling from YoYo attacks. Detecting YoYo attack allows cloud vendors to mitigate the attack by applying different strategies such as inline scrubbing or by holding cluster downscale of Nodes to maintain the short response time on the expense of high economic damage.
Unfortunately, we were unable to access real life data from Kubernetes production clusters that can supply big data to feed our model. Our research aims to analyze a live Kubernetes cluster reacting dynamically.  Considering that, although our analysis is focused on ML methods evaluation and results high accuracy detecting the YoYo attack, we can not guarantee at this point the same results on a production Kubernetes cluster. 

Time series data generated by the Kubernetes cluster are the primary foundations of the features for our model (e.g., response time, pod count, node count and CPU load as you can see see in Figure \ref{figure:yoyo_attack}). The input data contains hundreds of thousands discreet features (e.g., each value recorded is considered a feature). We evaluated a deep neural network approach where a convolution neural network (CNN) was concatenated with long short term memory (LSTM) module which is good fit for a sequential data, followed by a linear layer to classify a YoYo attack. This deep neural network requires a large amount of data and long training cycles, this method is not optimal to sparse datasets. In order to match our unique input data to machine learning methods we used a feature extraction principles. We found out that extracting statistical functions from the time series data (e.g mean, minimum, maximum,std,median) generating up to 20 features per sample is ideal for our model to achieve excellent performance. We evaluated multiple machine learning methods and selected the XGBoost \cite{10.1145/2939672.2939785} algorithm. XGBoost is a non linear classifier that works well for our attack detection evaluation. 
\subsection{Model Setup and Formulation}
\label{sec:yoyo_detector}
\subsubsection{XGBoost Classification Algorithm}
 XGBoost is a state of the art method that addresses the best speed and accuracy with limited datasets. XGBoost is a classifier and a regressor based on the gradient boosting algorithm. It uses an ensemble of decision trees. Each tree corrects the mistakes of the trees that came before it. It is both fast and efficient, performing well on sparse data and on a wide range of predictive modeling tasks.\\
 
We explain the XGBoost detection method \cite{8367124} in a nut-shell. $R^m$ is the space of  regression trees. Let q represent the structure of each tree that maps a sample to a corresponding leaf index. Let T be the number of leaves in the tree.  Each $f_k$ corresponds to an independent tree structure $q$ and leaf weights $w$. Each regression tree contains a continuous score on each leaf. The final prediction for a given sample is the sum of predictions from each tree. 
The tree ensemble model is trained in an additive mode and it greedily sums up the gradients on each leaf and then applies the scoring value to improve the target function, and brings the result to a minimum. 
\ignore{
{The target function of the XGBoost model is deﬁned in equations \ref{eq:xgboost_loss_1} to \ref{eq:xgboost_loss_3}:}
\begin{eqnarray}
     \label {eq:xgboost_loss_1}
     F_{Obj}(\Theta) = L(\Theta)+ \Omega(\Theta) \\
     L(\Theta) =l(\widehat{y_i},yi) ,\\
     \label {eq:xgboost_loss_2}
     \Omega(\Theta)= \gamma{T} + \frac{1}{2}\lambda ||W||^2
     \label {eq:xgboost_loss_3}
\end{eqnarray}   
The objective function consists of two parts:  $L(\Theta)$ and $\Omega(\Theta)$. Theta refers to the various parameters in the formula. Among them, $L(\Theta)$ is a difference-convex loss function that measures the difference between the prediction $\widehat{y_i}$ and the target ${y_i}$. The function indicates how to ﬁt the data for our model. Commonly used convex loss functions such as mean square loss function and logistic loss can be used in the above equation.  $\Omega(\Theta)$ is a regularized term that penalizes complex models. The key parameters to optimize our prediction are the number of leaves in the tree (T), the learning rate $\gamma$ whose value is between 0 and 1, and the maximum depth. The additional regularization term $\gamma$, when multiplied by T, is equal to the number of edges that were pruned in order to obtain the spanning tree. This helps to avoid over-ﬁtting. The algorithm is very powerful since it enumerates over all possible splitting points greedily. However, it is impossible to  do so efficiently when the data does not fit entirely in memory. Therefore, the algorithm behaves as an approximation algorithm: it proposes candidate splitting points according to percentiles of feature distribution. The algorithm then maps the continuous features into buckets split by these candidate points, aggregates the statistics and finds the best solution among proposals based on the aggregated statistical results.}
\subsubsection{Methodology \& Data Engineering}
Our test-bed design for the YoYo attack on Kubernetes is documented in detail in Section 4 (Experiment analysis). We built our datasets on a live cluster, experimenting on labeled traffic loads with the following two classes: \emph{Attack} (1) and \emph{Regular} (0). Class label \emph{Attack} (1) represents a YoYo attack as described above, and class label \emph{Regular} (0) represents an average load on a site. Testing and training datasets are a collection of balanced experiments of the two classes taken with a range of parameters as describe in \ref{tbl:parameters }. Note, the default setting for the on-attack time to $t_{on} = 10$ minutes and off-attack time to $t_{off} = 20$ minutes. The default settings for the YoYo attack is $k=20$ and for \emph{Regular} load is $k=2$.  The collection is split randomly with 70\%:30\% ratio for training and testing datasets. We are interested in learning the cluster autoscaling behavior when the system is under attack and when the system is normally or highly but legitimate loaded. The goal is to binary classify the situation. We mimicked normal and attack distributions loads by setting HTTP connections using the popular JMeter v5.2 \cite{Jmeter} network tool. Hence we started by generating a workload that simulates YoYo attacks. To increase the dataset variance we populated multiple parameters and values. We trained the model with multiple $t_{on}$ and $t_{off}$ attacks, covering in each attack or regular sample at least 3 cycles of duration $T$. Thus, our model requires at least 3 of cycles of duration T for inference. We configured threads ramp-up time (The ramp-up period defines the duration the full number of threads are loading) to control load scale-up in different levels. Last, we used JMeter timers to set either a constant or random delay between requests in the \emph{Regular} class.  We believe that using all of these parameters simulate similar conditions as best as possible of a real distributed load on typical web applications. Table \ref{tbl:parameters } represents the variance of datasets created using experiments parameters.
\begin {table}[ht]
\caption{ Dataset parameters for YoYo classifier} \label{tbl:parameters } \centering
\vskip 2pt
\resizebox{\linewidth}{!}
{
\begin{tabular}[t]{|l||c|c|c|c|}
\hline
\normalsize & \normalsize ramp-up(sec)  & \normalsize $t_{on}$/$t_{off}$(min) & \normalsize Power(k) & \normalsize timers \\   
\hline
\hline
\normalsize Regular & \normalsize 30,60,120 & \normalsize continuous   & \normalsize 1,3,5,7 & \normalsize constant,random\\
\hline
\normalsize Attack & \normalsize 30,60,120 & \normalsize 7/14, 10/20, 12/24 & \normalsize  15,20,30 & \normalsize constant \\
\hline
\end{tabular}
}
\end {table}

\subsubsection{XGBoost Optimization}
We used Python Scikit-learn model selection interfaces such as RandomizedSearchCV and GridSearchCV to optimized the XGboost parameters. The parameters of these estimators used to optimized our selected method parameters by cross-validated search over parameter settings and can apply scoring. The parameters we optimized using these methods are: number of estimator trees (The more trees defined, the better are the estimates and longer is the computation time), max depth (parameter to control over-fitting as higher depth will allow model to learn relations very specific to a particular sample. The absolute maximum depth would be $N-1$, where N is the number of training samples), max features and criterion (Impurity function: gini vs entropy). We found by using cross validation estimator the following parameters values as the best to achieve the highest performance: $class weight= balanced$, $criterion= gini$, $max\_depth= 1$, $max\_features= auto$, $min\_{samples}\_{leaf}= 10$, ${min\_samples\_split}= 40$, ${n\_estimators}= 10$. In addition, we executed the Explainable Boosting Machine package (EBM) to understand the most explainable features for the classifications among the 20 selected features. Table \ref{tbl:ebm} ranks each feature with a value between 0 and 10 where 0 is the least important feature for the model and 10 is the most important feature.
 \begin {table}[ht]
\caption{Overall importance absolute score for XGBoost features} \label{tbl:ebm} \centering
\vskip 2pt
\resizebox{\linewidth}{!}
{
\begin{tabular}[t]{|l||c|c|c|c|c|}
\hline
\normalsize Feature name & \normalsize Mean  & \normalsize Std  & \normalsize Maximum & \normalsize  Minimum & \normalsize Median \\
\hline
\hline
\normalsize Response Time & $6.0$ & $6.0$ & $8.8$ & $0.0$ & $7.8$\\
\hline
 Pods & 9.3 & 9.8 & 9.5 & 0.0 & 7.2\\
\hline
\normalsize CPU load & 4.4 & 4.5 & 10.0 & 0.0 & 1.0\\
\hline
\normalsize Nodes & 9.2 &  9.2 &  9.8 & 5.5 & 9.3\\
\hline
\end{tabular}
}
\end {table}
\subsection {Evaluation Results} 
In this section we summarize the results of our models as they perform on our dataset. We collected experimental data from 21 samples of attacks and the same regular load as described above. We provide the results of classiﬁcation measures for accuracy, precision, recall, F1, false positives (FP), false negatives (FN), true positives (TP) and true negatives (TN).\\
\begin{equation}
     \label {eq:models_metric}
accuracy= \frac{TP + TN}{TP + FP + TN + FN} 
\end{equation}
\begin{equation}
    precision = \frac{TP}{TP + FP}
\end{equation}
\begin{equation}
    recall = \frac{TP}{TP + FN} 
\end{equation}
\begin{equation}
    F1 = 2\cdot\frac{precision \cdot recall}{precision + recall}
\end{equation}
  
\subsubsection {XGBoost Classifier Evaluation } 
 The XGBoost classifier has the most accurate results on our Datasets as you can see in table \ref{tbl:algos}. We evaluated multiple machine learning methods in addition to XGboost. Among them are: Logistic regression, Random forest, Decision trees and a deep neural network based on CNN + LSTM. XGBoost model is a perfect fit for experiments with either sparse data, it requires less performance time (no epochs are required), and it suffers the least from over-fitting. 
\ignore{ 
\begin {table}[ht]
\caption{ YoYo XGBoost classifier results on our dataset} \label{tbl:xgboost} \centering
\vskip 2pt
\resizebox{\linewidth}{!}
{
\begin{tabular}[t]{|l|c|c|c|c|c|c|c|c|}
\hline
\normalsize & \normalsize TP & \normalsize TN & \normalsize FP & \normalsize FN & \normalsize precision & \normalsize recall & \normalsize F1 & \normalsize accuracy\\   
\hline
\normalsize Training &\normalsize 16 & \normalsize 15 & \normalsize 0 & \normalsize 1 &\normalsize 1.0 & \normalsize 0.94 & \normalsize 0.97 & \normalsize $96\%$\\
\hline
\normalsize Testing & \normalsize 9 & \normalsize 8 & \normalsize 1 & \normalsize 0 & \normalsize 0.89 & \normalsize 1.0 & \normalsize 0.94 & \normalsize $95\%$\\
\hline
\end{tabular}}

\end {table} 
Table \ref{tbl:xgboost} shows a detail split of train and test datasets results.}

Our proposed classifier can achieve the highest accuracy score on the testing dataset, the XGBoost algorithm reaches an accuracy score of 95\% (F1=0.94), while the CNN+LSTM, Logistic Regression, Decision Tree and Random Forest can achieve accuracy score which is less than 90\%.  The running time of XGBoost and other classic machine learning algorithms is less than a second. The running time of CNN+LSTM is much higher,it is counted in thousands seconds due to the enormous number of epochs. Therefore, overall, our selected model has the best comprehensive performance.
\begin {table}[ht]
\caption{ Algorithms performance comparison.} \label{tbl:algos} \centering
\vskip 2pt
\resizebox{\linewidth}{!}
{
\begin{tabular}[t]{|l||c|c|c|c|}
\hline
\normalsize & \normalsize Recall & \normalsize Precision & \normalsize  Accuracy \% & \normalsize Training Time(sec) \\   
\hline
\hline
\normalsize XGBoost & \normalsize 1.0 & \normalsize 0.89 & \normalsize 94\% & \normalsize 0.33 \\
\hline
\normalsize CNN+LSTM &\normalsize 0.85 & \normalsize 1.0 & \normalsize 93\% & \normalsize 3600  \\
\hline
\normalsize Random Forest &\normalsize 0.83 & \normalsize 0.82 & \normalsize 83\% & \normalsize 0.15  \\
\hline
\normalsize Logistic Regression &\normalsize 0.78 & \normalsize 0.78 & \normalsize 85\% & \normalsize 0.5  \\
\hline
\normalsize Decision Tree &\normalsize 0.83 & \normalsize 0.89 & \normalsize 84\% & \normalsize 0.04 \\
\hline
\end{tabular}}

\end {table} 
\section{\uppercase{Related Work}} \label{sec:relatedWork}
DDoS prevention is crucial to cloud computing environments \cite{5487489}. Several works \cite{amazonDDoSwhitepapers,miao2014nimbus} recommend auto-scaling and Kubernetes specifically as a possible solution to mitigate DDoS attacks.  Some works \cite{vivinsandar2012economic,somani2015ddos} describe how a traditional DDoS attack can be transformed into an EDoS in the cloud environment. Other works have tried to mitigate EDoS attacks \cite{8291616,9036257,8460025,8367124} using machine learning classification techniques trying to limit a malicious bot. A recent work uses the XGBoost classifier as a DDoS detection method \cite{8367124} in an SDN based cloud.  
Other studies \cite{8064125} research containerized autoscaling but  refer neither to cloud resilience nor DDoS mitigation.
\ignore{ emerging costly premium services are offered by few companies  such as Reblaze{\texttrademark} for full DDoS protection \cite{Reblaze}, covering all layers of networking. Other solutions implemented  predictive mechanism in the cloud provider aiming to detect attackers that perform burst attacks using anomaly detection techniques.}A recent work \cite{8460025} focused on resource patterns with cyclic trends with the aim of analyzing resource behavior on micro-services using auto-aggressive statistical models on auto-scaling systems. Older works attempt to prevent EDoS by directing suspicious traffic to a scrubber service and using client puzzles to detect legitimate users by identifying suspicious traffic using a threshold policy (e.g., requests/sec)\cite{6375171}.

\section{\uppercase{Conclusions}} \label{sec:conclusion}
In this work we illuminate the potential of exploiting the auto-scaling mechanism to perform an efficient attack on Kubernetes that impacts the cost and the quality of a service. We show that Kubernetes is still vulnerable even with a light-weight containerized architecture. We also show that YoYo VM attack results almost the same relative economic damage as the YoYo Kubernetes attack. However, VM groups are still less resilient to YoYo attacks than Kubernetes clusters. In addition we conclude that the YoYo attack on Kubernetes is more cost effective for the attacker than a classic DDoS attack. We believe that the auto-scaling mechanism alone is not enough, and therefore we propose a unique approach based on XGBoost algorithm to detect YoYo attacks and allow to mitigate the DDoS damage.\ignore{by analyzing unique Kubernetes patterns in addition to the corresponding traffic.} We also show that XGBoost algorithm has an high accuracy and a lower false positive rate.\ignore{ XGBoost is a scalable method that fits well to the complex Kubernetes autoscaling platform.} To the best of our knowledge this work is the first to detect DDoS burst attacks on a Kubernetes cluster using a machine learning method, specifically with XGBoost.

Future works will aim to evaluate YoYo attacks on multiple applications from different tenants running in the same Kubernetes cluster when only one of the applications is the target. Future research may also evaluate the resilience of cloud mesh services running multi-tenant environments to YoYo and similar DDoS attacks.
\ignore{
We concluded that the best auto-scaling option that an admin should choose as the best choice mitigation for DDoS and EDoS attack is to select a machine type with $R$ high enough to load large enough number of Pods per Node. Defining Nodes wisely for the cluster will utilize Pods without launching extra Nodes.  }

\section*{Acknowledgments}
This research was supported by Google Cloud Research. We would like to thank also Daniel Bachar and Assaf Sinvani for helpful discussions and thier help with the experiments setup.
\bibliographystyle{plain}
\bibliography {references.bib}

\end{document}